\documentstyle[aps,preprint,amssymb]{revtex}
 \tightenlines
\begin{document}
\draft
\title{Quenched Disorder Effects on Deterministic
Inertia Ratchets}
\author{C.M. Arizmendi$^{1,2}$, Fereydoon
Family$^1$, A. L. Salas-Brito$^{1,3}$} 

\address {$^1$Department of Physics, Emory
University, Atlanta, GA 30322,  USA\\
$^2$Depto. de F\'{\i}sica, Facultad de
Ingenier\'{\i}a, Universidad Nacional de Mar del
Plata,\\  Av. J.B. Justo 4302, 7600 Mar del Plata,
Argentina\\
$^3$Laboratorio de Sistemas Din\'amicos,
Departamento de Ciencias B\'asicas,\\ Universidad
Aut\'onoma Metropolitana-Azcapotzalco, Apartado
Postal 21-726,\\  Coyoac\'an 04000 D.\ F., M\'exico}
\date{\today}
\maketitle
\begin{abstract} The effect of quenched disorder on
the underdamped motion of a periodically driven
particle on a ratchet potential is studied.  As a
consequence of disorder, current reversal and
chaotic diffusion may take place on regular
trajectories. On the other hand, on some chaotic
trajectories disorder induces regular motion. 
 A localization effect similar to {\sl Golosov
Phenomenon} sets in whenever a disorder threshold
that depends on the mass of the particle is
reached.  Possible applications of the localization
phenomenon are discussed.
\end{abstract}
\bigskip

Thermal ratchets \cite{general}   are simple
stochastic models where a nonzero  net drift  speed
may be obtained from time correlated fluctuations
interacting with  asymmetric periodic structures. 
The study of ratchets has received much attention
due to their  general interest in modeling molecular
motors \cite{motor}.  Another sources of interest is
the possible application of ratchets for modeling
nanoscale friction
\cite{friction}, the potential for the development
of new approaches for  separation
 of microscopic and mesoscopic objects
\cite{astumian}, models for understanding surface
smoothening
\cite{barabasi},  and the building of micron-scale
devices
\cite{build}.

 In thermal ratchets, thermal fluctuations are
rectified in different ways
 according to the type of ratchet system. For
example, in the {\sl rocking 
 ratchet}, which is the most common type of ratchet,
a time-dependent external driving force of zero
average acts as  the rectifier pumping mechanism. In
this kind of ratchet thermal noise  does indeed 
help the ratchet by increasing its
efficiency\cite{Takagi}.  In contrast,  spatial
disorder  reduces the characteristic drift speed
\cite{Marchesoni,Harms} of thermal ratchets. 

    Recently, the influence of quenched disorder, in
the absence of noise, on a 
    periodically forced overdamped 
    particle in a periodic asymmetric
    potential  was considered \cite{we}. An  outcome
of this study
 was the discovery of diffusive transport in the
presence of quenched disorder. Diffusion was
observed even with  small amounts of added disorder
and the diffusion current was found to increase with
increasing noise and eventually  reach the same
order of magnitude as the regular drift.  As it is
common in the study of thermal ratchets,  this
study  was carried out in the  Smoluchowski limit of
vanishing mass. However, inertial effects are
important in many experimental situations.  For
example,  the finite mass of the particles  plays an
important role in friction at the nanoscale
 as well as in microscopic particle separation
experiments.
 Thus, it would be of interest to study the effects
of quenched disorder on the dynamics of underdamped
thermal ratchets with finite mass.

  Inertial ratchets, even in the absence of noise,
have a very complex dynamics, including chaotic
motion  
 \cite{Jung,Mateos}.  This deterministically induced
chaos mimics, to some extent, 
 the role
 of noise \cite{chaos}.    This added complexity 
drastically changes some of the basic properties of
thermal ratchets.  For example, it has been shown
that  inertial ratchets can exhibit multiple
reversals in the current direction
\cite{Jung,Mateos}.  It has been suggested
\cite{Mateos} that this behavior may be related to
crisis in which a chaotic ratchet state suddenly
becomes periodic, but it was shown later
\cite{Barbi} that current reversals can occur even
in the absence of bifurcations from chaotic to
periodic motion.
 
 The aim of the present paper is to study the
effects of spatial disorder on the dynamics of the
underdamped deterministic ratchet, specially the
influence of disorder on regular and  chaotic motion
and  current reversals. We will concentrate on the
model of a  particle of non-vanishing mass,
periodically driven in an asymmetric periodic
potential with quenched disorder. No temporal noise
term is considered, just the quenched disorder.
 
In scaled non-dimensional coordinates the equation
of motion is  given by \cite{Jung}

\begin{equation}
\epsilon {{d^2 x} \over {dt^2}} + \gamma {{d x}
\over {d t}} = \cos (x) +
\mu \cos (2x) + \Gamma
\sin(\omega t) + \alpha~\xi(x). 
\label{motion} 
\end{equation}

\noindent Here, $\epsilon$ is the mass of the
particle, $\gamma$ is the damping coefficient,
$\Gamma$ and
$\omega$ are, respectively, the amplitude and
frequency of an external oscillatory forcing, and
$\alpha~\xi(x)$ is the
 term due to the quenched disorder.  The terms 
$\xi(x) $ are independent, uniformly distributed
random variables with no spatial correlations,
corresponding to a piecewise constant force on the
period of the potential.  The coefficient $\alpha
\geq 0$ is the strength of the quenched disorder.  
The unperturbed ratchet potential,
 
\begin{equation} 
     U(x) = - \sin(x) - \mu \sin(2x)
\label{potential} 
\end{equation} 

\noindent has been the subject of extensive recent
studies
\cite{we,Jung,Bartussek,Lindner} mainly in models
with no disorder. A recent work \cite {Mateos} has
analysed the influence of the chaotic behavior in
Eq.\ (\ref{motion}) with $\alpha=0$ and has related
it, with the help of a bifurcation diagram, to the
observed reversals in flow direction but an {\sl a
posterior} study has  reexamined some of its
conclusions
\cite{Barbi}. Here we consider the addition of
quenched disorder in order to analyze the effect of
a more  realistic representation of the substrate,
on the dynamics of the ratchet. We note that with
the disorder term included, the ratchet equation (1)
can  also be used  to model fluctuations in DC
current amplitude in arrays of Josephson junctions
\cite{josephson} and in studies of friction,
particularly the sliding motion of clusters on
surfaces \cite{friction}. 

Previous work \cite{Jung} has shown that in the
absence of quenched disorder ($\alpha = 0$) there
are both regular and chaotic solutions of Eq.\
(\ref{motion}) as well as multiple current
reversals---arguably related to crisis in the
underlying dynamics \cite{Mateos,Barbi}. In the
present work, we  study the influence of quenched
disorder on the system for both kinds of
trajectories (regular and chaotic).  Specifically,
numerical solutions of Eq.\ (\ref{motion}) were
obtained using a  variable step Runge-Kutta-Fehlberg
method
\cite{NumRec}.  We let $\epsilon=20$, $\gamma =
1.0$, $\mu=0.25$,
$\omega = 0.1$,  and  studied the behavior for
several values of $\Gamma$ (see below). The
calculational details are the same as in \cite{we}.

 In figure 1 we show a typical periodic trajectory
at $\Gamma=0.9245$ in the absence of disorder (the
thick line) and the corresponding trajectory with a
very small amount of quenched noise, $\alpha=0.001$
(the thin line).  It is apparent from the figure
that the  trajectory gets modified. Figure 2 shows
the corresponding phase portrait confined to the
$x$-interval $(-2\pi, 2\pi]$.   Notice that we
exhibit in the same plot both the case with no
disorder (the  points at the center of the six
squares) and the chaotic attractor it becomes after
the quenched noise is added. The phase portrait
confined to the same $x$-interval is valid with
disorder because the small disorder term may be
considered as a small perturbation.

In order to get a  global picture of the behavior,
in Figure 3 we show the bifurcation diagram of Eq.\
\ref{motion} as a function of $\Gamma \in [0.65,
1)$, both in the case with no disorder (3a) and with
a small,   
 $\alpha=0.01$, quantity of quenched noise added
(3b). The chosen range for $\Gamma$
  corresponds to the existence of regular and
chaotic solutions and inversion of
  current in the absence of disorder.  In all the
chaotic cases analysed, the sign of the current for
a given $\Gamma$
 coincides with the sign of the majority of $x$ of
the corresponding trajectory
 in the bifurcation diagram; for the periodic
states, on the other hand, we could not find any
definitive correlation between the current  and the
bifurcation behavior.

Figure 3b also shows the different ways in which
deterministic states are affected when  quenched
disorder is introduced. For example,
 for
$\Gamma
\in (0.8,0.9)$ (see also Fig. 2) states that are
periodic in the deterministic case are changed to
chaotic states.  Consider  a periodic state within
a  thin  zone in the bifurcation diagram (as in the
$\Gamma$-zone roughly between $0.854$ and
$0.864$ in figure 3a). The effect of the quenched
noise on the system
 is to send the ratchet to a nearby chaotic zone, as
it should be clear from  figures 3a and 3b.

On the other hand, some irregular states in the
region around
$\Gamma
\simeq 0.73$ of Fig. 3 are changed to more regular
states. This last effect can be associated with the
taming of chaos with disorder
\cite{yuri} where disorder induces ordered motion
characterized by very complex but nevertheless
regular patterns.  In a wide periodic window (like
the one roughly between $0.693$ and $0.722$) the
quenched noise has little effect on the periodic
behavior, except near its upper end where  the
chaotic zone itself becomes wider and superposes
itself on the originally regular zone. Such behavior
can also be seen by comparing figure 3a with figure
3b. The main point then is that the ratchet dynamics
follows the dominant behavior (of the $\Gamma$-zone
it belongs to) when a small  amount of quenched
noise is added. 

In a recent paper, Popescu et al.\ \cite{we} have 
shown that quenched disorder induces a significant
additional chaotic ``diffusive'' motion on the
overdamped version of Eq.\ (\ref{motion}). Thus, in
the finite inertia case we are considering, strong
fluctuations are expected and this calls for the use
of a time dependent probability measure, as has been
previously 
 done in \cite{we,Jung}.
 
A Gaussian distribution was chosen as the initial
probability density. Apart from early transients, a
linear mean and a linear variance, whcih are
characteristics of a Brownian motion, are observed.
The third  and higher  order cumulants increase 
 slower than $t^{n/2}$, up to $n = 6$, with $n$ the
order of the cumulant.
 Therefore, the probability distribution $p_t(x)$ is
asymptotically a Gaussian,
 and, as is well known, 
  the first  and second cumulants, associated with
the mean and the variance are
  sufficient to describe the asymptotic evolution of
$p_t(x)$
   \cite{we,Jung}.

Averages  were performed over ensembles of 5000
trajectories starting from different initial
conditions  very close to the origin $x = 0$. The
ensemble described above  was left to evolve  for
800 external drive periods $T$, and  every 10
periods the positions $x(t)$ were stored for
further  analysis.

We first consider the case of periodic behavior with
$\Gamma=0.9245$.  In Figures 4a and 4b  we show
results for the first and second  moments, {\sl i.e.}
$C_1(t) =
\langle  x(t)
\rangle$ and   
$C_2(t) = \langle {(x(t)-\langle x(t) \rangle)}^2
\rangle$, respectively,  where
$\langle~\dots~\rangle$ means average over the
ensemble, as a function of  time $t$ without
quenched disorder ($\alpha=0$), and with two
different small amounts of quenched disorder
($\alpha = 0.005$, and
$\alpha = 0.01$). For $\alpha=0$, $C_1(t) \simeq
v(\alpha)~t$ and
$C_2(t)$ tends to a constant as $t \rightarrow
\infty$ which characterizes a periodic state. For
$\alpha = 0.005$, and
$\alpha = 0.01$, both first and second moments show
an asymptotic linear dependence on time $t$, $C_1(t)
\simeq v(\alpha)~t$, 
$C_2(t) \simeq D(\alpha)~t$. There is a reversal of
current due to the presence of quenched disorder,
and at the same time the onset of diffusion. This
last effect is also obtained by adding a small
amount of quenched disorder to the overdamped
ratchet \cite{we}. 

To study the effect of quenched disorder on a
chaotic trajectory we now consider the case of 
$\Gamma=0.8967$.  We calculate the first and second
moments, $C_1(t)$ and
$C_2(t)$, both with no quenched disorder
($\alpha=0$) and with  different small amounts of
quenched disorder ($\alpha = 0.005$,
$\alpha = 0.01$ and $\alpha = 0.05$). The
corresponding figures 5a and 5b show that again
there is a current reversal and the magnitude of the
current  increases with increasing quenched
disorder. The existing diffusive behavior is also
slightly increased.\par

It is interesting to note that the reversal of the
current is not always associated with the addition
of quenched disorder, as can be seen in figures 6a
and 6b, where we have plotted
$C_1(t)$ and $C_2(t)$  for the chaotic trajectory
corresponding to
$\Gamma = 0.8955$ without quenched disorder
($\alpha=0$), and with  different  amounts of
quenched disorder ($\alpha = 0.03$, $\alpha =
0.05$,  and $\alpha_t \simeq 0.1$). The current
maintains its direction and increases with
increasing disorder until $\alpha=0.1$.  With
further increase in $\alpha$, the current  goes to
zero asymptotically.  Diffusion also increases with
increasing disorder for
$\alpha<0.1$, but it  tends to zero asymptotically
with increasing
$\alpha$. This localization effect may be clearly
seen in the particle trajectories when the particle 
gets stuck or oscillates with the same amplitude 
for several periods.
     
A localization effect in random walks on random
environments  is known as the {\sl Golosov
Phenomenon} in the random walk literature
\cite{Bouchaud}. Its ocurrence has  been proven
rigorously for systems with only nearest neighbor
transitions by Golosov \cite{golosov}. It was also
reported by G\"unter Radons \cite{Radons} on
one-dimensional chaotic maps where chaotic diffusion
is totally suppressed by the presence of quenched
disorder. The {\sl Golosov Phenomenon} may be
described as a packet of initially close particles
moving in a coherent fashion from one minimum to the
next deeper minimum. Hence, it may also occur  in a
disordered ratchet when and if particle motion
becomes locked to the external driving frequency.
Our results show that for all values of 
$\Gamma$ studied, current and diffusion
asymptotically tend to zero if the amount of
quenched disorder exceeds a certain treshold. It is
important to note that according to our preliminary
results,
 this threshold depends on the mass of the particle,
it being smaller when the mass of 
 the particle
  decreases. This is clearly an instance of a
localization effect analogous to {\sl Golosov
Phenomenon}. 

An important application of the localization effect
may be in
 microscopic particle separation \cite{astumian}. 
The reason is that
  the threshold value of quenched disorder for the
appearance of localization effect seems to depend on
the mass of the  particle. Thus, it may be possible
to develop a new approach for separating microscopic
particles without having to provide an applied
gradient.  This can be accomplished by tuning the
amount of quenched disorder on a microfabricated
sieve in a way that 
 particles of smaller mass, with lower disorder
threshold,
 remain stacked while   more massive particles and
with higher disorder threshold display a net current.

 In summary we have studied the effects  of small
amounts  of quenched disorder on an underdamped
(inertial) ratchet. In analogy with the case of
overdamped ratchets, we found that strong diffusive
motion may be induced in the periodic trajectories. 
On the other hand, we found that for some values of
$\Gamma$ associated with chaotic trajectories in the
overdamped case, quenched disorder induces regular
solutions. 
 Disorder can cause current reversals in both
chaotic and regular solutions, and whenever the
amount of quenched disorder exceeds a certain
threshold the motion becomes localized.  

 These results should be helpful in the
interpretation of  experimental results in studies
of friction, particularly at  the nanoscale, as well
as in understanding transport processes  in
molecular motors.
 An interesting direction for future research may be
to study the dependence of the  localization
phenomenon on the particle mass.  This question is
of interest  due to its possible application in
designing  new particle separation techniques.

%**************************************************************

\bigskip 
\noindent {\bf Acknowledgements}

This work was supported by grants from the Office of
Naval  Research, from the Universidad Nacional de
Mar del Plata, from the CONACyT, and from  the
Universidad Autonoma Metropolitana-Azcapotzalco. We
thank M.N. Popescu for very useful discussions.
C.M.A. acknowledges enjoyable discussions (not just
on ratchets) with George Hentschel and George A.
Bohling.  A.L.S.-B. acknowledges both the
hospitality and the support  of the Department of
Physics of Emory University and the leave of absence
granted by the Depto.\ de Ciencias B\'asicas UAM-A;
last but not least, he dedicates this work to his
late friends B. Ch. Caro, R. Micifuz, F. C. Cucho, 
and C. Ch. Ujaya.

\begin{figure}
\caption{Trajectory of the particle for
$\Gamma=0.9245$,
$\epsilon=20$, $\gamma = 1.0$, $\mu=0.25$, $\omega =
0.1$ with no quenched disorder (thick line) and with
a very small amount, $\alpha=0.001$, of quenched
disorder (thin line).}
\end{figure}

\begin {figure}
\caption{Phase portrait of two attractors of the
ratchet equation for
$\Gamma=0.9245$, $\epsilon=20$, $\gamma = 1.0$,
$\mu=0.25$, $\omega = 0.1$. With no quenched noise
the attractor consists of the points at the center
of the squares. With a very small amount,
$\alpha=0.001$, of quenched disorder the attractor
becomes chaotic. }
\end{figure}

\begin{figure}
\caption{a) Bifurcation diagram as a function of
$\Gamma$ for
$\epsilon=20$, $\gamma = 1.0$, $\mu=0.25$, $\omega =
0.1$ in the no quenched disorder case. 
b)Bifurcation diagram as a function of $\Gamma$ for
$\epsilon=20$, $\gamma = 1.0$, $\mu=0.25$, $\omega =
0.1$ with an amount, $\alpha=0.001$, of quenched
disorder.} 
\end{figure}

\begin{figure}
\caption{Influence of disorder on the probability
distribution for
$\Gamma=0.9245$, regular with no disorder for three
different amounts of quenched disorder;
$\alpha=0$, $\alpha=0.005$, and $\alpha=0.01$: a)
First moment $C_1(t)$ for 
$\epsilon=20$, $\gamma = 1.0$, $\mu=0.25$, $\omega =
0.1$ as a function of time. b) Second moment
$C_2(t)$ for
 $\epsilon=20$, $\gamma = 1.0$, $\mu=0.25$, $\omega
= 0.1$ as a function of time.}
\end{figure}

\begin{figure}
\caption{Influence of disorder on the probability
distribution for
$\Gamma=0.8967$, chaotic with no disorder for four
different amounts of quenched disorder; $\alpha=0$,
$\alpha=0.005$, $\alpha=0.01$ and $\alpha=0.05$:
a)First moment $C_1(t)$ for 
$\epsilon=20$, $\gamma = 1.0$, $\mu=0.25$, $\omega =
0.1$ as a function of time.b) Second moment $C_2(t)$
for  $\epsilon=20$, $\gamma = 1.0$,
$\mu=0.25$, $\omega = 0.1$ as a function of time.}
\end{figure}

\begin{figure}
\caption{Influence of disorder on the probability
distribution for
$\Gamma=0.8955$,  chaotic with no disorder.
Diffusion and current increase keeping direction of
current  until localization sets in. Four different 
amounts of quenched disorder are shown: $\alpha=0$,
$\alpha=0.03$, $\alpha=0.05$ and $\alpha=0.10$. At
$\alpha=0.1$ current and diffusion  vanish
asymptotically which means localization.
 a)First moment $C_1(t)$ for 
$\epsilon=20$, $\gamma = 1.0$, $\mu=0.25$, $\omega =
0.1$ as a function of time. b) Second moment
$C_2(t)$ for
$\epsilon=20$, $\gamma = 1.0$, $\mu=0.25$, $\omega =
0.1$ as a function of time.}
\end{figure}

\end{document}